\documentclass[a4paper, 10pt, conference]{ieeeconf}      

\IEEEoverridecommandlockouts                              

\usepackage{hyperref}
\usepackage{graphicx}
\usepackage{amsmath}
\usepackage{amssymb}
\usepackage{xcolor}
\usepackage{float}
\usepackage{subcaption}
\let\labelindent\relax
\usepackage{enumitem}

\IfFileExists{ulem.sty}{\usepackage[normalem]{ulem}}{\providecommand{\sout}[1]{##1}}
\definecolor{addblue}{RGB}{0,70,200}
\definecolor{delred}{RGB}{200,0,0}
\definecolor{phgray}{RGB}{128,128,128}
\newif\ifmarkup
\markuptrue
\ifmarkup
  
  \newcommand{\del}[1]{{\scriptsize\textcolor{delred}{\sout{#1}}}}
  
\else
  
  \newcommand{\del}[1]{}
  
\fi

\title{\LARGE \bf Location-Independent Robot-Assisted Finishing Using Digital Twins and Extended Reality}

\author{Jose Outeiro$^{1}$, Jia Holt$^{1}$, Tero Kaarlela$^{2}$, and Khalil Chakal$^{2}$ 
\thanks{$^{1}$Digital Engineering for Advanced Manufacturing Laboratory (DEAM Lab), Center for Precision Metrology, Department of Mechanical Engineering and Engineering Science, University of North Carolina at Charlotte, NC 28223, USA
        {\tt\small jc.outeiro@charlotte.edu}}%
\thanks{$^{2}$Materials and Mechanical Engineering, Faculty of Technology, University of Oulu,
        Oulu 90014, Finland
       }%
}

\begin{document}

\maketitle
\thispagestyle{empty}
\pagestyle{empty}

\begin{abstract}

This paper presents a cyber-physical system (CPS) for location-independent programming, supervision, training, and teleoperation of a Robot-Assisted Finishing (RAF) system used to post-process metal additive-manufactured (AM) components. A digital twin (DT) built in Unity is delivered to the operator as a WebGL application that supports both desktop and immersive modes through WebXR-compatible devices. Moreover, it exchanges robot state and pose commands with a collaborative robot through a Message Queuing Telemetry Transport (MQTT) broker. The DT enforces kinematic and collision constraints before a pose is released to the physical robot, and augments the virtual component with a color map of the surface topography that supports operator decisions on part repositioning or process termination. The architecture was validated on a specially designed physical RAF system. A steady-state joint synchronization error of $0.12^\circ$ and a mean round-trip latency of 563 ms were measured, which is adequate for supervisory programming and intermittent teleoperation.

\end{abstract}

\section{INTRODUCTION}
\label{sec:introduction}

Additive manufacturing (AM) produces components from a three-dimensional model by adding material layer by layer~\cite{Bigliardi2024}. AM has the potential for profitable small-batch and lot-size one production, with reduced material waste, shorter design-to-product time, and decentralized close-to-customer services, although the extent of these benefits depends strongly on part geometry, material, and batch size~\cite{Bigliardi2024}.

\subsection{Motivation}
\label{subsec:motivation}

For the geometries and batch sizes in which AM is competitive, production itself is highly automated, but the finishing step that follows remains largely manual and time-consuming.
To increase the overall AM profitability, automated and flexible methods are required for the finishing process of AM components~\cite{Krajnik2021}.

AM finishing processes involve noise, vibration, dust, and limited visibility into the finishing operation. Using the manufacturing system described in Section~\ref{subsec:physicaltwin}, a sound pressure level of about 85~dB(A) was recorded during finishing. While operators are required for supervision and process decisions, continuous physical presence near the finishing cell is not desirable.

Robotics is widely used for production tasks such as welding, assembly, and handling, but small-batch AM production requires flexible robot control methods that can be operated by personnel with limited robotics expertise, thereby reducing process time and manual labor.

Developing fully autonomous part finishing systems for small production batches and lot-size-one AM components may not be economically feasible. Human operators remain valuable because they can adapt finishing strategies to component geometry and process behavior. The proposed cyber-physical system (CPS) combines human decision-making and remote teleoperation of manufacturing systems via a Digital Twin (DT) using eXtended Reality (XR) as a human-robot interface, enabling economically viable finishing of low-volume production while reducing operator exposure to harsh and noisy environments.

\subsection{Scientific Contribution and Research Questions}
\label{subsec:scientificcontribution}

The proposed work extends previous educational DT frameworks toward industrial deployment, where DTs are utilized not only for training but also for robot programming, teleoperation, and training. The presented approach uses XR as the high-level human-robot interface and the Message Queuing Telemetry Transport (MQTT) communication protocol to provide location-independent supervision and control of the finishing process.

The present work builds on the authors' Cyber-Physical Machine Tool (CPMT) concept for CNC machining~\cite{Kaarlela2025cpmt} and a DT/XR battery-disassembly system~\cite{Kaarlela2025digital}. Like those systems, it uses a Unity DT delivered as a browser-hosted WebGL runtime and coupled to a physical asset through a cloud MQTT broker. Here, the controlled asset is a six-degree-of-freedom collaborative robot, requiring inverse kinematics and whole-arm collision checking; the DT produces reusable robot programs and carries process-level information in addition to kinematic state.

The main contributions of this work are:

\begin{enumerate}[leftmargin=1.4em]
    \item A human-in-the-loop CPS architecture for a RAF system remote teleoperation, including the safety and cybersecurity measures required to control a manufacturing system over a public network.
    \item An XR-enabled DT framework supporting location-independent robot programming, teleoperation, and training.
    \item A simplified offline robot programming system using the DT and XR human-robot interface, and synchronization with the physical robot to reduce production downtime.
    \item A DT-based process-awareness visualization enabling human remote monitoring and control of the finishing process.
    \item A CPS for workforce training and education in automation/robotics and manufacturing processes.
\end{enumerate}

The work addresses two research questions:
\begin{enumerate}
    \item \textit{RQ1: Is the latency and synchronization performance achievable with a browser-hosted DT and a cloud MQTT broker sufficient for the remote teleoperation of a RAF system?}
    \item \textit{RQ2: What process information must the DT carry, beyond the robot kinematic state, for an operator to monitor and control the RAF system remotely?}
\end{enumerate}

\section{STATE-OF-THE-ART}
\label{sec:relatedwork}

\subsection{Digital twins and cyber-physical systems}

Grieves introduced the DT as a conceptual model for product lifecycle management~\cite{grieves2016}. Kritzinger \textit{et al.} distinguish a Digital Model, which has no automated data exchange; a Digital Shadow, which has automated flow from the physical to the digital entity only; and a Digital Twin, which has automated bidirectional flow~\cite{kritzinger2018}. This taxonomy is applied to the present system in Section~\ref{subsec:classification}.

\subsection{Industrial robot programming}

Recording waypoints with a hand-held teach pendant has been the standard method of programming industrial robots for decades~\cite{zhang2020}. It requires expertise in the specific robot model and the operator's physical presence, which is undesirable in noisy environments, and programming complex components with many waypoints is time-consuming. Hand-guiding removes the need for programming skills but still requires physical presence and interruption of production. Offline programming and CAD/CAM-driven path generation remove the presence requirement, but for lot-size-one AM components the path must be adapted to each part, and the decision of when a surface is sufficiently finished depends on in-process observation rather than on a fixed program. This motivates an approach in which the operator remains in the loop but is not physically present.

\subsection{Extended reality for robot programming and teleoperation}

XR human-machine(robot) interfaces for industrial manipulators are an active field. Solanes \textit{et al.} teleoperated industrial manipulators through augmented reality~\cite{Solanes2020}. Togias \textit{et al.} designed robot paths in a virtual reality environment~\cite{Togias2021}. Garg \textit{et al.} programmed and simulated industrial robots in VR using a DT~\cite{Garg2021}. Shu \textit{et al.} presented a platform-independent programming interface~\cite{Shu2022platform}. Rofatulhaq \textit{et al.} built a virtual engineering platform for online learning, reporting MQTT communication latencies of the order of tens of milliseconds~\cite{Rofatulhag2020}. These systems are evaluated on generic pick-and-place or path-following tasks and do not carry information about the manufacturing process being performed.

\subsection{Robot-assisted abrasive finishing}

Centrifugal disk finishing is an established abrasive finishing process for improving the surface finish of as-built metal AM parts. Kopp and Uhlmann correlated simulated contact quantities with measured roughness reduction on AM workpieces and predicted \emph{local} differences in roughness reduction on surfaces with limited media accessibility~\cite{Kopp2022}. Thus, contact count alone is not a valid proxy for material removal.

\subsection{Research gap}
Table~\ref{tab:relatedwork} summarizes the reviewed systems against the capabilities targeted in this work. No reviewed system combines location-independent teleoperation of a manufacturing system, an immersive cross-platform interface, and process-level feedback specific to the proposed RAF system.

\begin{table}[htbp]
\centering
\caption{Comparison with reviewed systems. TO: teleoperation of a physical system; LI: location-independent; IM: immersive; XP: cross-platform; PA: process-awareness feedback.}
\label{tab:relatedwork}
\begin{tabular}{l c c c c c}
\hline
Work & TO & LI & IM & XP & PA \\
\hline
Garg \textit{et al.}~\cite{Garg2021}        & Yes & No  & Yes & No  & No \\
Togias \textit{et al.}~\cite{Togias2021}    & Yes & No  & Yes & No  & No \\
Solanes \textit{et al.}~\cite{Solanes2020}  & Yes & No  & Yes & No  & No \\
Shu \textit{et al.}~\cite{Shu2022platform}  & Yes & Yes & No  & Yes & No \\
Rofatulhaq \textit{et al.}~\cite{Rofatulhag2020} & Yes & Yes & No & Yes & No \\
Kaarlela and Outeiro~\cite{Kaarlela2025cpmt} & Yes & Yes & Yes & Yes & No \\
Kaarlela \textit{et al.}~\cite{Kaarlela2025digital} & Yes & Yes & Yes & Yes & No \\
Kopp and Uhlmann~\cite{Kopp2022}            & No  & No  & No  & No  & Yes \\
\textbf{This work}                          & Yes & Yes & Yes & Yes & Yes \\
\hline
\end{tabular}
\end{table}

\section{SYSTEM ARCHITECTURE AND IMPLEMENTATION}
\label{sec:proposedconcept}

\subsection{Physical manufacturing system}
\label{subsec:physicaltwin}
\begin{figure}[!tb]
  \centering
  \includegraphics[width=0.60\columnwidth]{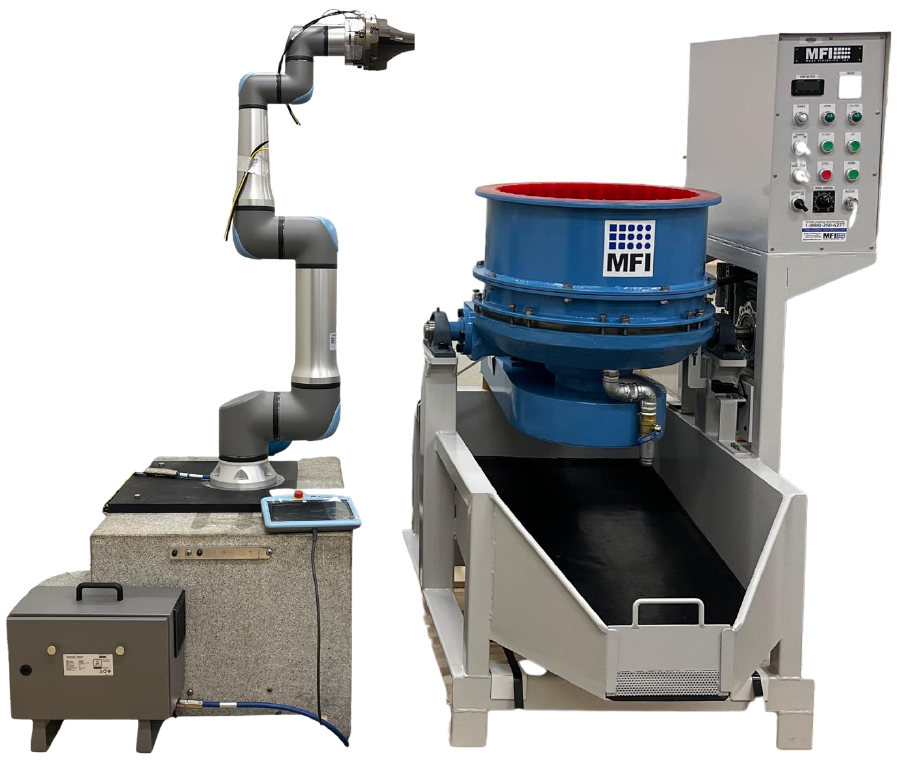}
  \caption{RAF system composed by a Universal Robots, model UR30, and a centrifugal disk finisher from Mass Finishing, model RF-50.}
  \label{fig:RAF_system}
\end{figure}

The physical manufacturing system is a RAF system for finishing AM components. It comprises a collaborative robot from Universal Robots, model UR30, and a centrifugal disk finisher from Mass Finishing, model RF-50. The robot has six degrees of freedom (six-dof), a reach of 1.3 m, and a payload of 30 kg. It is installed on a granite pedestal and is equipped with a three-finger, pneumatically actuated gripper. The finisher consists of a centrifugal bowl with a tilt mechanism to facilitate the discharge of the finishing media. The RAF system is illustrated in Figure~\ref{fig:RAF_system}, and is installed at the Department of Mechanical Engineering and Engineering Science at the University of North Carolina at Charlotte. 


\subsection{Digital Twin}
\label{subsec:digitaltwin1}

The virtual user interface is built using the Unity game engine to enable a realistic environment for the robot teleoperator.
A three-dimensional CAD model of the RAF system was assembled in SolidWorks, with the dimensions of the centrifugal disc finisher taken from the manufacturer's technical documents and the UR30 model downloaded from the manufacturer website. The robot's components and kinematics are based on the ROS\_industrial package~\cite{rosuniversal2024}. The model was textured in Blender and exported to Unity, where articulated joints connect the robot's joints and define the angular constraints of each joint, and where the bidirectional data flow between the DT and the physical system was implemented.
The user-interface functionality is implemented as a C\# script attached to an overlay game object. It provides the functions common to arm robots: joint and linear jog controls that inch the robot in a chosen direction at a preset velocity, gripper open and close, and recording, simulation, and playback of programmed waypoints.

In addition to robot teleoperation, the DT enables virtual viewpoints that are difficult or impossible to obtain in the physical RAF system. The teleoperator can inspect the robot, the component, and the finishing bowl from arbitrary viewpoints, improving awareness during the finishing process without interrupting production. Furthermore, the finishing bowl walls can be disabled to expose the finishing media and the part to the human operator.

\subsection{Classification of the system}
\label{subsec:classification}

Following the taxonomy of Kritzinger \textit{et al.}~\cite{kritzinger2018}, the robot layer of the system is a full DT: joint and Cartesian states flow automatically from the physical robot to the DT, and validated pose commands flow automatically back. The centrifugal bowl, in contrast, is not yet fully instrumented. No force, vibration, or acoustic signals are currently acquired from the bowl or the part, so the DT only represents the RAF system kinematics and not the physics of the interaction between the abrasive media and the part. The only process data transferred to the DT is the surface topography at specific intervals of time. Therefore, fully instrumenting the process layer is the primary item for future work identified in Section~\ref{sec:conclusions}.

\subsection{Kinematics and pose validation}
\label{subsec:kinematics}

From the robot perspective, the DT converts operator-specified Cartesian motions into joint positions with an iterative damped least-squares (DLS) solver~\cite{Wampler1986}, which computes the joint increment given by,

\begin{equation}
\Delta \boldsymbol{q} = \boldsymbol{J}^{\mathsf{T}}\left(\boldsymbol{J}\boldsymbol{J}^{\mathsf{T}} + \lambda^{2}\boldsymbol{I}\right)^{-1} \Delta \boldsymbol{x} ,
\label{eq:dls}
\end{equation}

\noindent where $\boldsymbol{J}$ is the manipulator Jacobian mapping joint velocity space to end-effector velocity space \cite{Craig2005}, $\Delta \boldsymbol{x}$ is the Cartesian error, and the damping factor $\lambda$ trades tracking accuracy against conditioning near singularities. A damped iterative solver was preferred over the closed-form solution because it degrades gracefully at singular and out-of-reach targets, which the operator can request freely in an unconstrained XR interface. Kinematic joint limits are enforced through Unity articulation joint objects. Collision elements are defined for the gripper, robot arm, pedestal, and centrifugal bowl; the part and the finishing media are also included.



\subsection{Communication}
\label{subsec:communication}
MQTT is a lightweight publisher/subscriber messaging protocol widely used in Internet of Things applications, relying on a dedicated message broker and supporting authentication, authorization, and encryption~\cite{mqtt5}.
Communication latencies on the order of tens of milliseconds have been reported for MQTT in teleoperation applications~\cite{Rofatulhag2020}. As shown in Section~\ref{subsec:latency}, the end-to-end latency of the present system is dominated not by the broker but by the polling interval of the robot controller program. The DT publishes to the topic \texttt{raf/robot/cmd} and subscribes to \texttt{raf/robot/state}. The message schema is shown in Figure~\ref{fig:architecture2}. Each message carries a monotonically increasing sequence number and a publisher timestamp so that the subscriber can reject stale or out-of-order commands.


\subsection{Safety architecture}
\label{subsec:safety}

The DT validates poses before they are published, but it executes in the operator's web browser and is therefore outside the trust boundary of the RAF system: any client able to publish to the command topic can request motion. Safety is consequently enforced at two levels. In the DT, invalid poses cannot be recorded or transmitted, which prevents the operator from issuing them accidentally. In the robot controller, the loop program independently rejects commands falling outside the configured joint limits and Cartesian safety planes before applying them, so that a malformed, replayed, or spoofed message cannot move the robot outside the safeguarded volume.

The RAF system is configured under the safety requirements of ISO~10218-1 and ISO~10218-2, and the collaborative operation limits of ISO/TS~15066 apply during any period in which a person may be inside the safeguarded space~\cite{iso102181, iso15066}. Because the commanding operator is remote, a local person may enter the cell without knowledge of pending remote motion. The RAF system therefore uses a light curtain that inhibits remote command execution.

\subsection{Cybersecurity}
\label{subsec:cybersecurity}

The cloud server hosts the DT runtime over HTTPS and the MQTT broker on port 8883, with all other ports closed at the firewall. Both services accept only TLS-encrypted connections using a certificate issued by the University of North Carolina at Charlotte. The broker requires per-user authentication and restricts publication rights on the command topic to authenticated operator accounts. The robot controller connects outbound to the broker and accepts no inbound connections. The servers are subject to frequent vulnerability scanning and remediation.

\section{RESULTS AND DISCUSSION}
\label{sec:results}

\begin{figure}[!tb]
  \centering
  \includegraphics[width=\columnwidth]{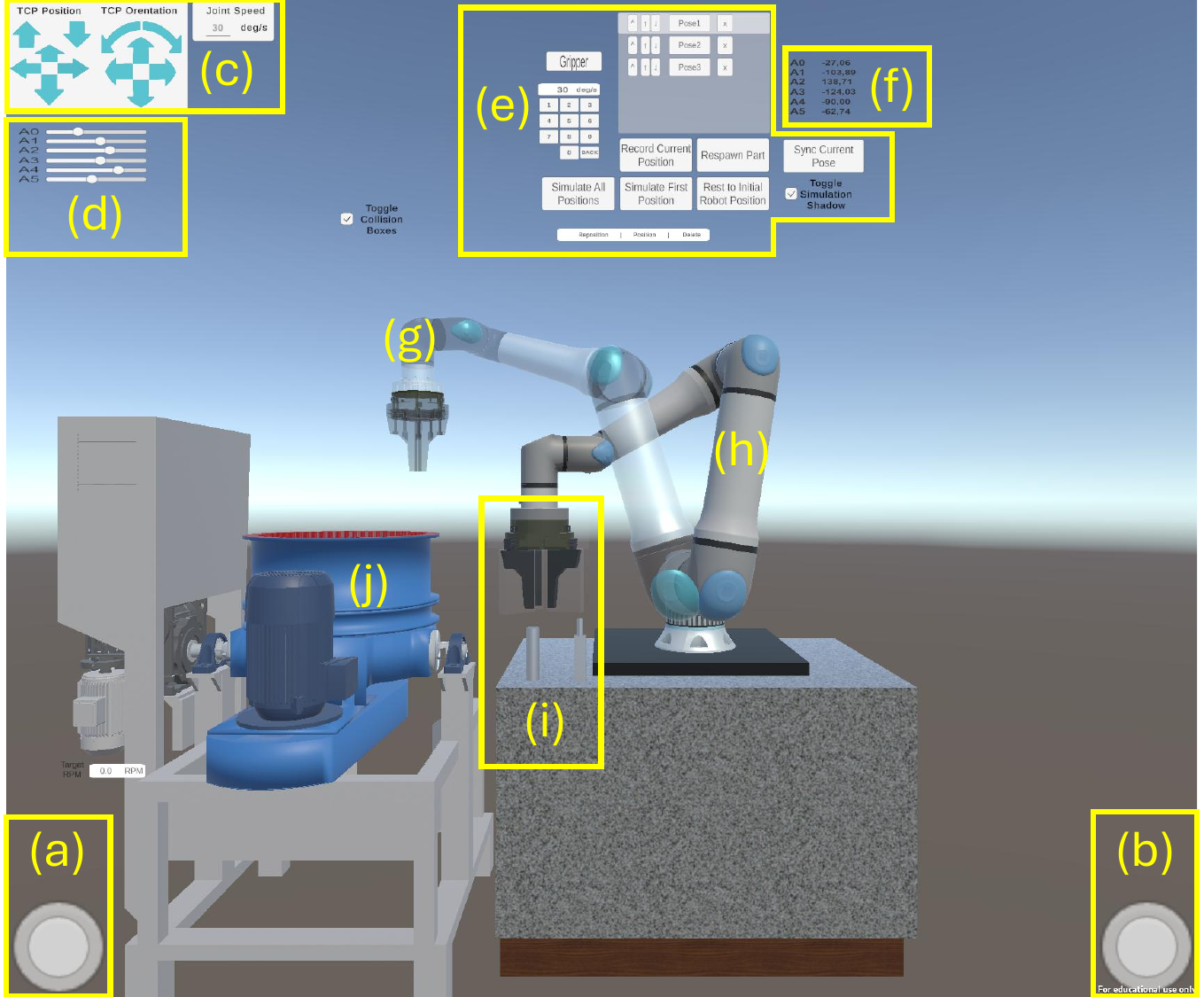}
  \caption{The virtual environment and user interface: (a) viewpoint pan, (b) viewpoint rotate, (c) robot TCP linear and angular control, (d) robot joint control, (e) pose record and program execution, and simulation controls, (f) physical twin joint positions, (g) shadow representing physical robot position, (h) DT, (i) gripper holding the AM part, (j) centrifugal bowl.}
  \label{fig:XRview}
\end{figure}

The main result is the DT of the RAF system, enabling teleoperation and monitoring of the physical system for programming, teleoperation and training. The user interface allows the operator to manipulate the robot arm, by grabbing and dragging arrow-shaped canvas elements with hand-held controllers, and to save robot poses using button elements, so that the kinematic and collision handling of Section~\ref{subsec:kinematics} is exposed directly to the operator. Figure~\ref{fig:XRview} illustrates the user interface and the canvas elements.

\subsection{Architecture}
\label{subsec:architecture}

\begin{figure*}[!tb]
  \centering
  \includegraphics[width=2\columnwidth]{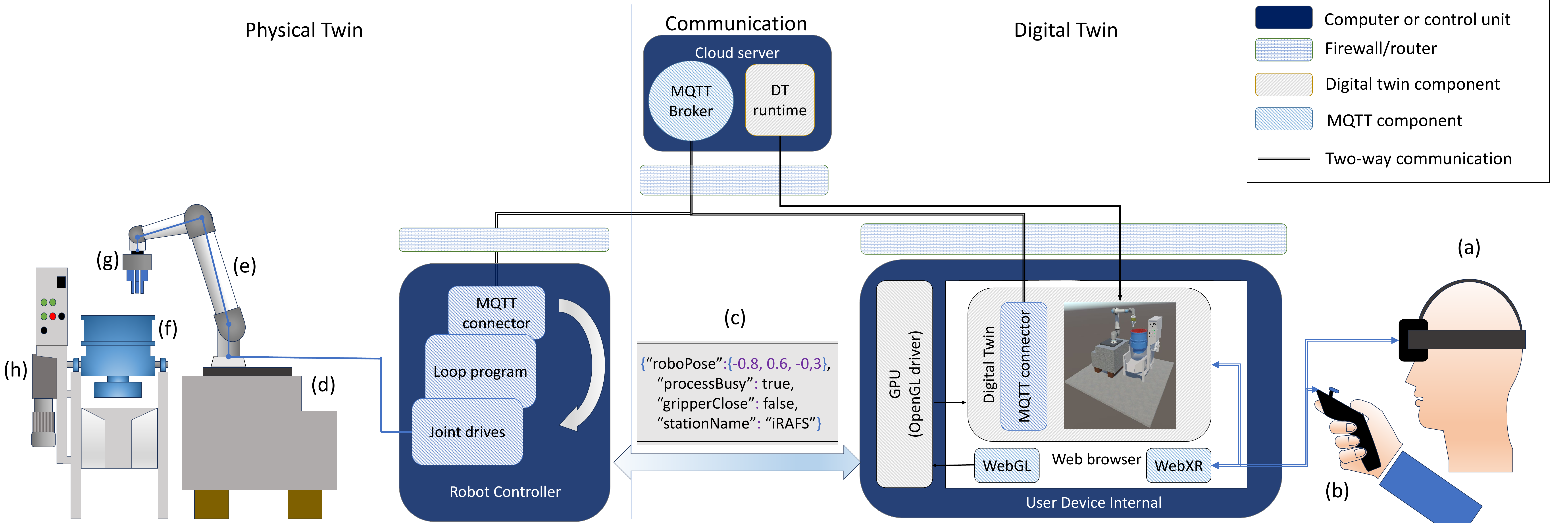}
  \caption{System architecture (a) VR/MR headset, (b) hand-held controllers (optional), (c) JSON message, (d) robot pedestal, (e) collaborative robot, (f) centrifugal bowl, (g) three-finger gripper, (h) tilt mechanism.}
  \label{fig:architecture2}
\end{figure*}

The cloud server architecture enables location-independent access to teleoperate the RAF system for monitoring and control: the WebGL-based DT runtime is hosted on the cloud server and is downloaded and executed by the user's web browser, which natively supports the compiled Unity web runtime~\cite{Craven2023}. The cloud server also hosts the MQTT broker, which enables bidirectional operational data updating between the DT and the RAF system. The architecture is illustrated in Figure~\ref{fig:architecture2}.

The Unity MQTT Bundle add-on~\cite{MQTTbundle2024} enables low-latency MQTT communications for the DT to teleoperate the physical system. The \textit{MQTT Connector Professional} third-party software by 4Each Software Solutions (4Each s.r.o., Praha, Czech Republic)~\cite{4Each2024} enables MQTT publishing and subscribing on the robot controller. The MQTT publisher periodically publishes the robot's actual joint and Cartesian positions, and the subscriber writes the requested joint or Cartesian position values to the robot controller.

The desired robot joint position and gripper state are updated in the robot controller's runtime, which runs a simple loop. The program polls the desired position and tool variables, updating the robot position or recalling a gripper close or open action.


\subsection{Process-awareness visualization}
\label{subsec:processawareness}

\begin{figure}[H]
  \centering
  \includegraphics[width=0.3\columnwidth]{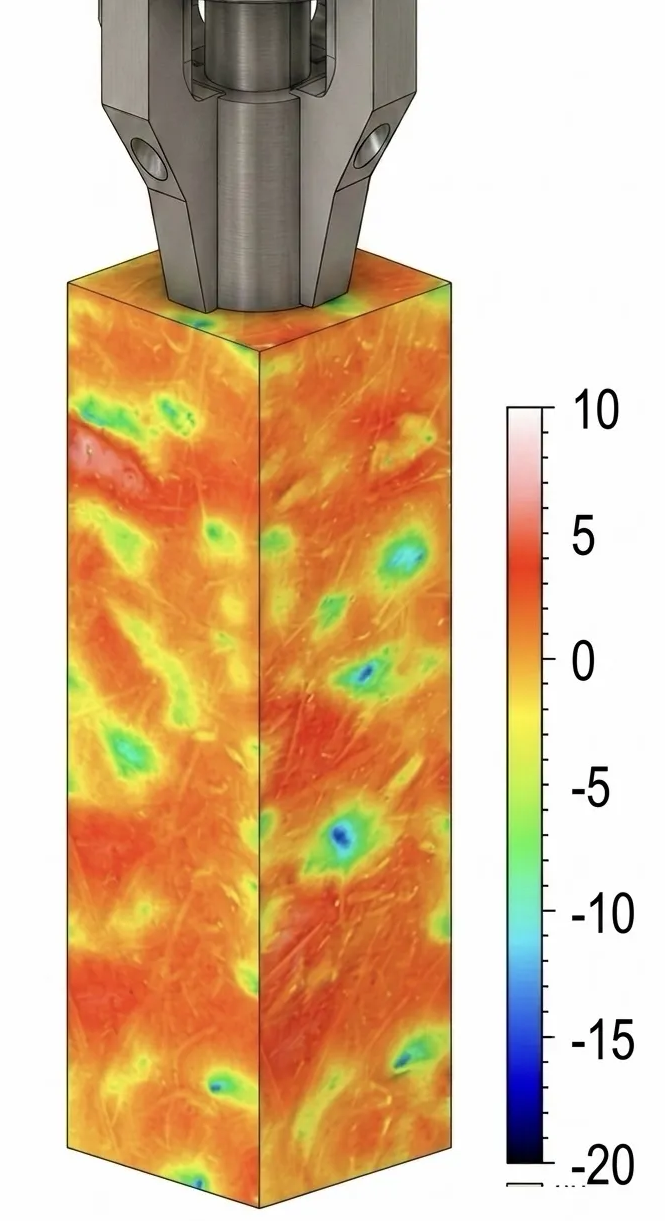}
  \caption{Height map representing the surface topography of the part during finishing.}
  \label{fig:colormap_roughness}
\end{figure}

Unlike conventional DTs of robotic systems, the developed DT also provides visual feedback related to the finishing process itself. The surface topography of the part is measured offline at fixed intervals of time by 3D Coherent Scanning Interferometry (CSI) using Zygo ZeGage system to determine its height map, aerial surface roughness and its uniformity. This information is used by the operator to decide whether to continue the process and change the position of the part in the media, or to terminate the process if the target surface roughness is reached. The visualization enables the teleoperator to identify areas receiving less finishing exposure and to reposition or rotate the component in the media accordingly. This functionality supports human decision-making during small-batch finishing operations where fully autonomous process planning may not be economically feasible. Figure~\ref{fig:colormap_roughness} illustrates the visual feedback of the part surface finishing.

\subsection{Validation}
\label{subsec:validation}

\begin{figure}[H]
    \centering
    \begin{subfigure}[t]{0.42\columnwidth}
        \centering
        \includegraphics[width=\linewidth]{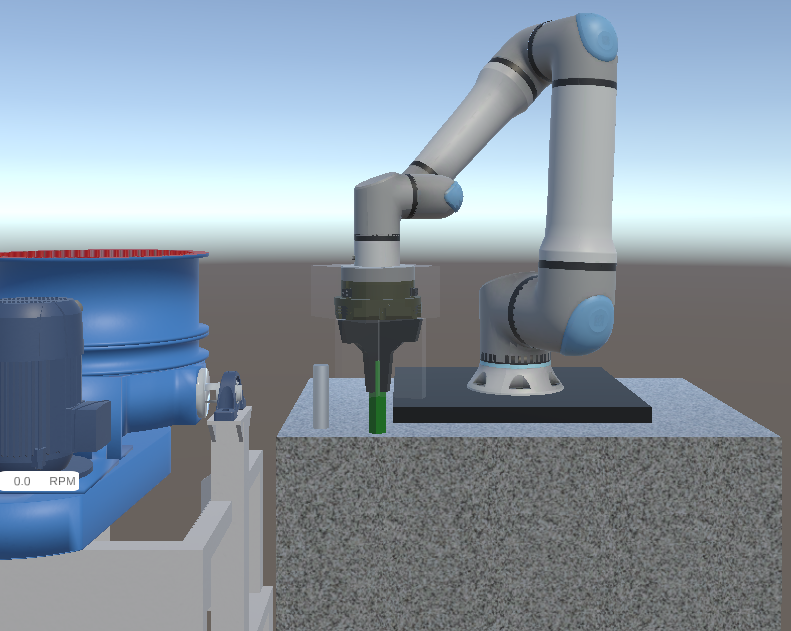}
        \caption{Approach and grip the part.}
        \label{fig:firstW}
    \end{subfigure}
    \hfill
    \begin{subfigure}[t]{0.42\columnwidth}
        \centering
        \includegraphics[width=\linewidth]{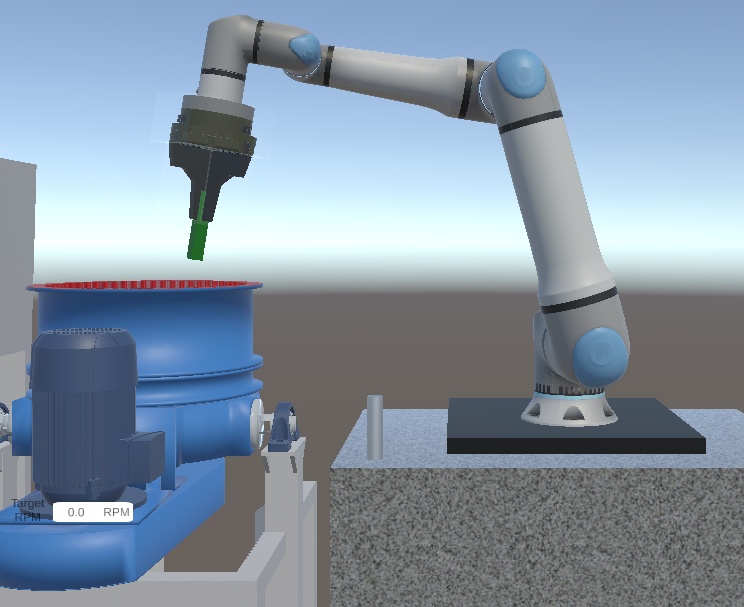}
        \caption{Position the robot above the bowl.}
        \label{fig:secondW}
    \end{subfigure}
    \vspace{0.5em}
    \begin{subfigure}[t]{0.42\columnwidth}
        \centering
        \includegraphics[width=\linewidth]{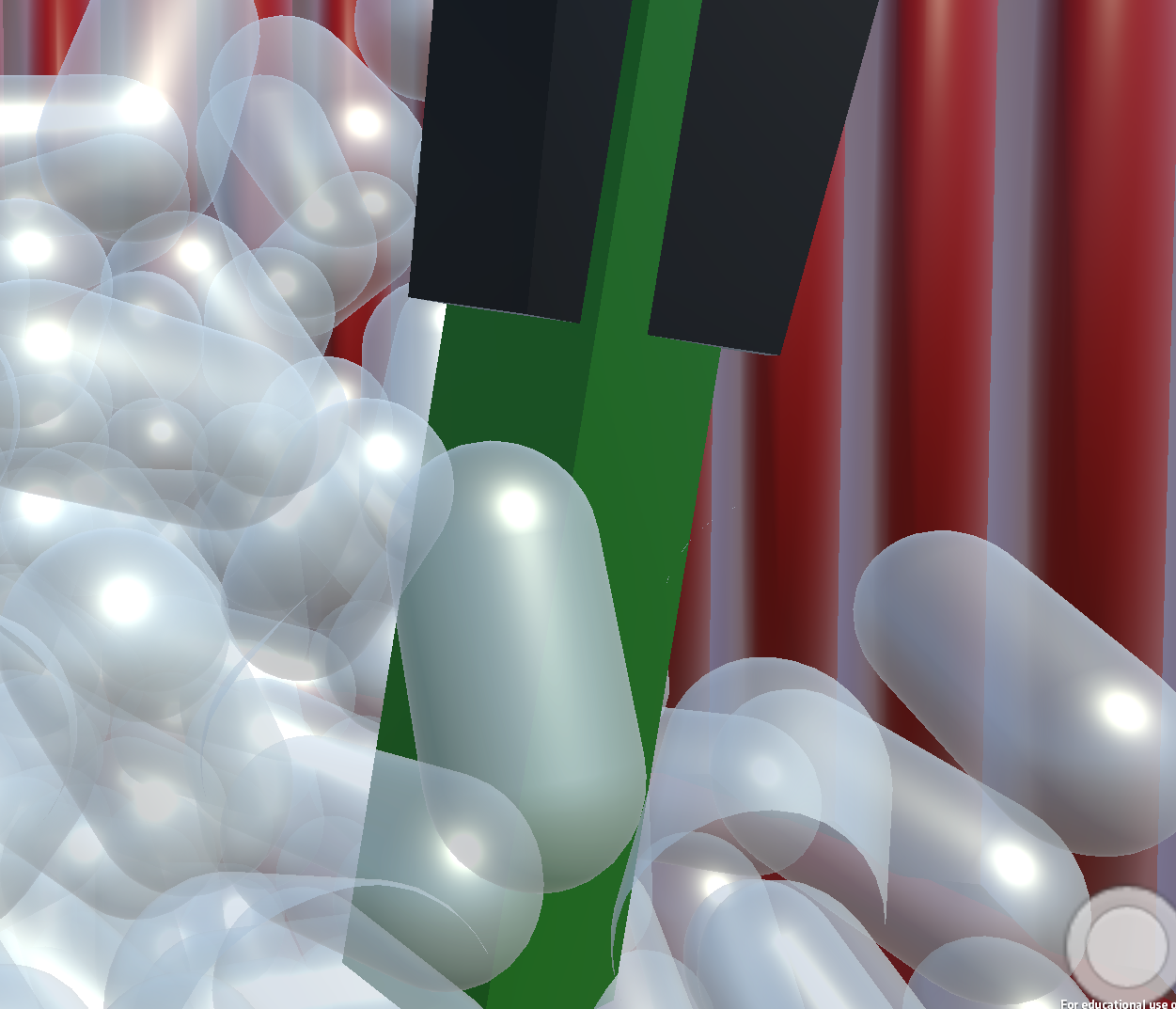}
        \caption{Position the part inside the bowl and reposition until finished.}
        \label{fig:thirdW}
    \end{subfigure}
    \hfill
    \begin{subfigure}[t]{0.42\columnwidth}
        \centering
        \includegraphics[width=\linewidth]{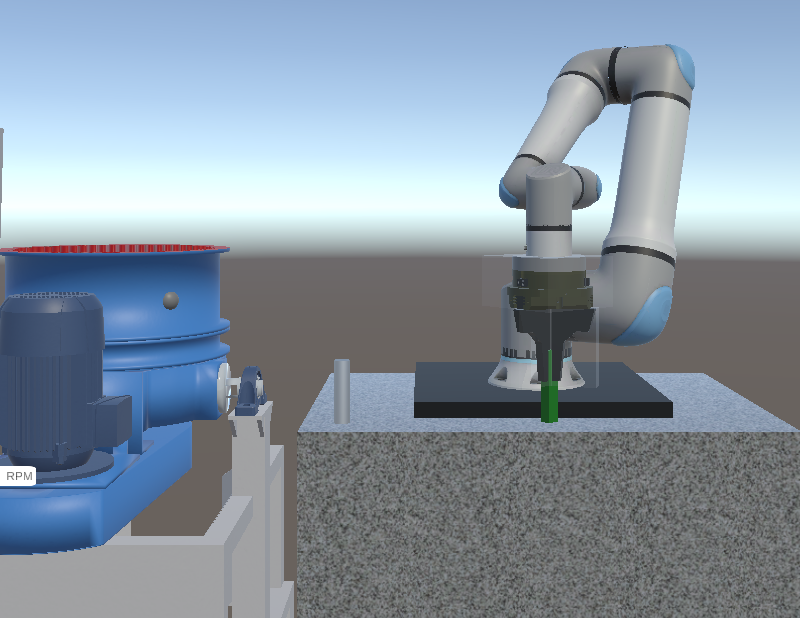}
        \caption{Remove the finished part from the bowl and place it on the table.}
        \label{fig:fourthW}
    \end{subfigure}
    \caption{Human-robot finishing workflow enabled by the proposed CPS.}
    \label{fig:workflow}
\end{figure}

A pick-and-place task was defined to validate the functionality of the proposed architecture: pick up the AM part from a table next to the robot, place it into the centrifugal bowl, and create a trajectory to finish the surface. This requires programming the robot to approach, pick, place, finish, retract, and control the gripper; all of these functions are integrated into the user interface, in which the teleoperator moves the robot model to the desired position and then initiates the DT. The robot positions can be programmed only if the DT validates the chosen end positions as collision-free. The finishing workflow is illustrated in Figure~\ref{fig:workflow}, showing how human decision-making and robotic execution are combined through the CPS.

\subsection{Performance}
\label{subsec:latency}

Table~\ref{tab:performance} shows the performance indicators of the proposed CPS. The maximum steady-state synchronization error of $0.12^\circ$ and the minimum and average round-trip latency are 431~ms and 563~ms, respectively.

\begin{table}[htbp]
\centering
\caption{Performance evaluation of the proposed CPS architecture}
\label{tab:performance}
\footnotesize
\begin{tabular}{p{5 cm} r}
\hline
Metric & Value \\
\hline
Joint sync.\ error, steady state, max. & $0.12^\circ$ \\
Joint sync.\ error, steady state, RMS & not reported \\
Round-trip latency, minimum & 431 ms \\
Round-trip latency, mean & 563 ms \\
Message loss rate & not reported \\
DT synchronization & Successful \\
Remote robot programming & Successful \\
Remote robot execution & Successful \\
\hline
\end{tabular}
\end{table}

The measured latencies exceed the 250~ms usability threshold adopted in the authors' earlier CPMT work~\cite{Kaarlela2025cpmt} and are an order of magnitude above the MQTT latencies reported for teleoperation in~\cite{Rofatulhag2020}, which indicates that the protocol is not the limiting factor. The dominant term is the polling interval of the controller loop program. The consequence is that the system does not support continuous motion streaming or direct manual control: at these latencies the operator is necessarily in a move-and-wait regime, the classical response to transmission delay in teleoperation~\cite{Ferrell1965}. This is compatible with the supervisory programming task addressed here, in which poses are recorded one at a time and verified in the DT before execution, but it precludes the reactive corrections an operator would make when guiding a tool into contact.


\subsection{Response to the research question}
\label{subsec:responseRQs}

Regarding RQ1, the achieved performance is sufficient for offline programming and intermittent teleoperation but insufficient for continuous teleoperation. An operator can compose, verify, and execute a finishing program remotely, but cannot react to an unexpected event during motion faster than the round-trip delay. Because the latency budget is dominated by the controller polling loop rather than by the network or the broker, the most effective improvement is to replace polling with an event-driven interface on the controller side.

Regarding RQ2, the robot's kinematic state alone is not sufficient for remote supervision of centrifugal disk finishing. Information from the process, including the surface topography, is required to determine the actions to be implemented, including the termination of the finishing process if the target surface roughness is reached.





\section{CONCLUSIONS AND FUTURE WORK}
\label{sec:conclusions}

The presented CPS enables economically viable finishing of small-batch and lot-size-one additive-manufactured components through a hybrid human-robot workflow that, instead of replacing the operator, combines human decision-making through an XR-enabled DT with the robotic execution of the physical finishing process.

The system enables location-independent programming, teleoperation, and operator training for the finishing process while reducing operator exposure to harsh working conditions. The DT provides process-awareness information through the visualization of surface topography, allowing operators to make informed repositioning decisions during finishing or to terminate the process.

The authors piloted the presented approach on a small variety of VR headsets and in desktop mode. Experimental validation demonstrated successful remote programming and execution of finishing tasks, with a maximum steady-state joint synchronization error of 0.12$^\circ$ and an average round-trip latency of 563 ms. The implementation is a laboratory prototype using industrial systems, not yet available for production use.

Three lines of future work are identified: (1) instrumenting the process with force, acoustic, and vibration sensors at the bowl and in-situ metrology equipment to measure the surface topography without removing the part from the gripper to reduce the total time to finish a part; (2) replacing the controller polling loop with an event-driven interface to reduce the dominant latency term; and (3) a user study with non-expert participants is required to substantiate the usability claims made for the interface.



\section*{ACKNOWLEDGMENT}

This research was supported by the University of North Carolina at Charlotte through the Center for Precision Metrology Affiliates Program, and the RIS4E - Revolutionary and Intelligent Steel Solutions for Sustainable Environment (Business Finland 43/31/2026).


\bibliographystyle{IEEEtran}
\bibliography{citations}

\end{document}